\documentclass[twocolumn,twoside,11pt]{elsart}

\usepackage{natbib}

\usepackage{graphicx}

\usepackage{amssymb}

\newcommand{\tend}{\mathrel{\mathop{- \kern -5pt - \kern -5pt
\longrightarrow}\limits_{n\to\infty}}}

\def\rmd {\kern 1pt {\rm d}}

\def\barR{\kern 2pt \overline{\kern -2pt R}}

\newcommand{\beeq}{\begin{equation}}
\newcommand{\eneq}{\end{equation}}
\newcommand{\beqn}{\begin{eqnarray}}
\newcommand{\eeqn}{\end{eqnarray}}

\def\mybig{\displaystyle \strut }

\def\dd{\partial}
\def\la{\raise.16ex\hbox{$\langle$}\lower.16ex\hbox{}  }
\def\ra{\, \raise.16ex\hbox{$\rangle$}\lower.16ex\hbox{} }
\def\go{\rightarrow}

\def\onehalf{ \hbox{${1\over 2}$} }

\def\eff{{\rm eff}}

\def\psibar{ \psi \kern-.65em\raise.6em\hbox{$-$} }

\def\myfrac#1#2{{\mybig #1\over \mybig #2}}

\begin{document}

\begin{frontmatter}


\title{Unified Theory of Elementary Particles \\
-- in Search of Extra Dimensions --}

\author{Yutaka Hosotani}
\address{Department of Physics,
Osaka University, Toyonaka, Osaka 560-0043, Japan}

\begin{abstract}
\rm
\quad Even though the unified theory of electroweak interactions is 
very successful at low energies, there remains  one part to be 
 confirmed.  It is the sector involving Higgs 
particles.  Those Higgs particles are expected to be discovered.  
It has been shown recently that Higgs 
particles can be viewed as gauge fields in higher dimensional gauge 
theory.  The mass of a Higgs particle and its coupling to other 
particles are constrained by the gauge principle.  In this scenario 
the mass of a Higgs particle is predicted to be in the range of  120 
GeV  -  290 GeV,  exactly in the range explored at LHC, provided that 
the extra dimension is curved and warped.  Thus 
physics of extra dimensions can manifest itself in collider 
experiments at the LHC energies. 
\end{abstract}

\end{frontmatter}

\section{Unification in extra dimensions}
At the most fundamental level quarks and leptons interact 
with each other by exchanging  gauge bosons.
Strong interactions are described by $SU(3)_C$ color gauge 
interactions whereas electroweak interactions by $SU(2)_L \times
U(1)_Y$ gauge interactions.  The associated gauge bosons are
gluons, $W$ bosons, $Z$ bosons, and photons.  There is one more 
field  necessary to make the standard model of elementary
particles to work.  It is the Higgs field which not only 
spontaneously breaks the electroweak symmetry $SU(2)_L \times
U(1)_Y$ to the electromagnetic symmetry $U(1)_{EM}$, but also 
gives fermions finite masses.  There appear many parameters
whose values are chosen to fit the observed data.  There is 
no principle regulating the Higgs sector of the standard model.

This seemingly awkward dilemma is resolved in higher dimensional 
gauge theory.  Long time ago Kaluza and Klein proposed an intriguing
scenario in which  we are living in five-dimensional spacetime.[1] 
They assumed that our spacetime is close to the 
product of four-dimensional Minkowski spacetime ($M_4$) and 
a circle ($S^1$) with
a radius $R$.  The metric in the five-dimensional space, 
$g_{MN}$ ($M,N = 0, \cdots, 4$), decomposes into the four-dimensional
metric $g_{\mu \nu}$ ($\mu , \nu = 0, \cdots, 3$),  the off-diagonal
components $g_{\mu 4}$, and $g_{44}$.  The general 
coordinate invariance in the fifth dimension implies that
the $g_{\mu 4}$ components behave as four-dimensional 
elecromagnetic gauge potential $A_\mu$.  In this manner the 
four-dimensional gravity and electromagnetism are unified in the 
five-dimensional gravity.

Motivated by Kaluza and Klein's idea, we consider non-Abelian 
gauge theory in five-dimensional spacetime.  Gauge potential 
decomposes into two parts;
\beeq
\quad
A_M = ( ~ A_\mu ~,~ A_y ~)  ~.
\label{decomposition1}
\eneq
On $M_4 \times S^1$, for instance, fields are expanded in Fourier
series in the fifth coordinate $y$;
\beeq
\quad
A_M(x,y) = \sum_{n=-\infty}^\infty A_M^{(n)} (x) ~ e^{iny/R} ~.
\label{fourier1}
\eneq
$A_\mu^{(0)} (x)$ describes four-dimensional gauge fields.
$A_y^{(0)} (x)$ tranforms as a four-dimensional scalar.   It is 
our contention that $A_y^{(0)} (x)$ contains the four-dimensional
Higgs scalar field.  Thus the Higgs field is a part of gauge 
fields and the unification of gauge fields and Higgs fields is
achieved.  The scenario is called the gauge-Higgs unification.[2,3]

\section{Dynamical gauge-Higgs unification}

To apply the gauge-Higgs unification scenario to the electroweak
interactions, several ingredients must be implemented.

\noindent
(i) In the electroweak theory $SU(2)_L \times
U(1)_Y$ breaks down to $U(1)_{EM}$ and the Higgs fields 
transform as a doublet of the $SU(2)_L$ group.  On the
other hand, the extra-dimensional component of the gauge fields
in the decomposition in (\ref{decomposition1}) belongs to the 
adjoint representation of the gauge group.  This means that one
must begin with a larger group to achieve gauge-Higgs unification,
as first clarified by Fairlie and by Manton.[2]

\noindent 
(ii) The electroweak symmetry is spontaneously broken by the 
4-d Higgs fields, which are a part of the 5-d gauge fields.
Dynamical electrowek symmetry breaking is induced by the 
Hosotani mechanism.[3,4]  When the extra-dimensional space is 
non-simply connected, there appear non-Abelian generalization of the 
Aharonov-Bohm phases.  Those non-Abelian Aharonov-Bohm phases, 
$\{ \theta_j \}$,  
become dynamical degrees of freedom, even though they give 
vanishing field strengths at the classical level.  At the 
quantum level those $\theta_j$, in general, develop nonvanishing
expectation values, thus breaking the gauge symmetry.

\noindent 
(iii) Quarks and leptons are chiral in the electroweak theory.
Left-handed and right-handed fermions interact with other 
particles differently.  The most natural and powerful way of
incorporating chiral fermions in higher dimensional theory is to
have an orbifold in extra dimensions.[5]  As a typical example,
consider $M_4 \times S^1$.  An orbifold is obtained by identifying
two points on $S^1$:
\vskip 3pt
\centerline{
$(x^\mu , y ) \sim (x^\mu, -y)$ .
}
\vskip 3pt
\noindent
The resultant spacetime is $M_4 \times (S^1/Z_2)$. 

\noindent 
(iv) As is seen below, phenomenology emerging from 
dynamical gauge-Higgs unification in flat space contradicts 
with the observation.  To have realistic phenomenology in
Higgs particles, quarks, and leptons, extra-dimensional space
should be curved.  In particular, dynamical gauge-Higgs unification in 
the Randall-Sundrum warped space yields intriguing consequences which
can be tested in the experiments at LHC.[6]

\section{Extra dimensions : flat or curved?}
One example in flat space with dynamical gauge-Higgs unification in 
electroweak interactions is given by the $U(3)_S \times U(3)_W$ model.[7]
There are two relevant non-Abelian AB phases ($\theta_1 , \theta_2$).
The effective potential $V_\eff(\theta_1, \theta_2)$ is depicted in fig.\ 1.
The abolute minimum is located at $(\theta_1, \theta_2) = (0, \pm 0.269 \pi)$
so that the electroweak symmetry $SU(2)_L \times U(1)_Y$ breaks down to
the electromagnetic symmetry $U(1)_{EM}$.

\begin{figure}[hbt]
\hskip .5cm
\includegraphics[width=6.cm]{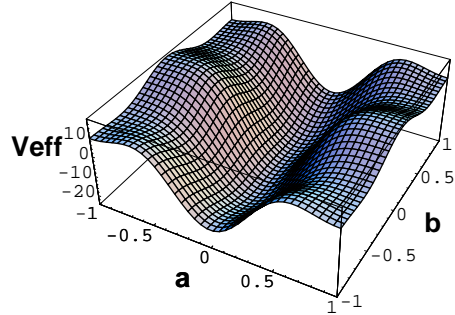}
\caption{The effective potential for the non-Abelian AB phases 
($\theta_1$=$\pi a \,$,$\, \theta_2$=$\pi b$) in the 
$U(3)_S \times U(3)_W$ model in flat space.}
\end{figure}

Although the symmetry is dynamically broken, there are two major problems.
The $W$ boson mass is predicted to be $0.135 /R$ where $R$ is the radius of 
extra-dimensins.  It implies that the Kaluza-Klein mass scale 
$M_{KK} = 1/R$ is about 600 GeV, which is too low.  The mass of the Higgs particle
is estimated from the curvature of the effective potential at its global
minimum.  One finds that $m_H \sim 0.871 \times \sqrt{\alpha_W} \, m_W$
where $\alpha_W = g_4^2/4\pi$ is the weak fine structure constant.
It leads to $m_H \sim 10 $ GeV, contradicting with experimental data.

These two features are generic in flat space.  The observational fact that 
the Higgs mass should be much bigger than $m_W$ indicates that the extra-dimensional
space, if it exsits, must be curved.

The most promising spacetime in the context of dynamical gauge-Higgs 
unification is the Randall-Sundrum (RS) warped spacetime.[8,9]  
It has the same
topology as $M_4 \times (S^1/Z_2)$.  The metric is given by
\beqn
\quad &&
ds^2 = e^{-2\sigma(y)} dx_\mu dx^\mu + dy^2 ~, \cr
\quad &&
\sigma(y) = k |y| \quad \hbox{for } |y| \le \pi R ~, \cr
\noalign{\kern 5pt}
\quad &&
\sigma(y+2\pi R) = \sigma (y) ~.
\label{RS1}
\eeqn
As in $M_4 \times (S^1/Z_2)$, $(x^\mu, y)$ and $(x^\mu, -y)$
are identified.
It is the anti-de Sitter space with a cosmological constant
$- 1/k^2$ sandwitched by two branes at $y=0$ and $y=\pi R$.
It has been speculated that five-dimensional anti-de Sitter 
space naturally emerges from such a more fundamental theory as
superstring theory.  Further reduction to four dimensions
yields approximately conformal theory with gauge fields and
light fermions.

The RS spacetime is specified with two parameters, $k$ and $R$.
It is natural to suppose that the structure of spacetime
is determined at the Planck scale $M_{pl} = 1.2 \times 10^{19}$ GeV.  
As a consequence it is expected that $k = O(M_{pl})$.  
The size $R$ is determined such that the theory predicts 
$m_W = 80.4$ GeV.  As is shown below, it implies that
$kR = 12 \pm 0.3$.

\section{$W$ bosons and the Kaluza-Klein mass}
Consider $SU(3)$ gauge group which contains $SU(2)_L \times U(1)_Y$.
Boundary conditions for the gauge fields in the RS spacetime  (\ref{RS1})
are given by
\beqn
&&\hskip -.2cm
\pmatrix{A_\mu \cr A_y} (x, y_j - y) 
= P_j \pmatrix{A_\mu \cr -A_y} (x, y_j + y) P_j^\dagger ~, \cr
\noalign{\kern 7pt}
&&\hskip 1cm
y_0 = 0 ~,~ y_1= \pi R ~, \cr
\noalign{\kern 3pt}
&&\hskip 1cm
P_0 = P_1 = \pmatrix{-1 \cr & -1 \cr &&+1} ~.
\label{BC1}
\eeqn
With these boundary conditions zero modes (massless modes) of 
$A_M$  in four dimensions exist only for components
\beeq
\quad
A_\mu = \pmatrix{ \circ & \circ \cr \circ & \circ \cr &&\circ}
~,~
A_y = \pmatrix{ && \circ \cr && \circ \cr \circ &\circ}
\label{zeromode1}
\eneq
where $\circ$ is marked. The zero modes of $A_\mu$ are $W$ bosons, $Z$ bosons,
and photons, whereas those of $A_y$ constitute the Higgs doublet.
In particular, the zero mode of $A_y$ gives rise to non-Abelian 
Aharonov-Bohm phase $\theta_W$:
\beqn
\quad &&
\exp \bigg\{ \myfrac{i}{2} \theta_W \Lambda \bigg\} = 
P \exp \bigg\{ ig \int_0^{\pi R}dy \, \la A_y \ra \bigg\} ~, \cr
\quad &&
\la A_y \ra  = c e^{2ky} ~\Lambda ~~,~~ 
\Lambda = \pmatrix{\cr &&1\cr &1} ~.
\label{ABphase1}
\eeqn

When $\theta_W \not= 0$ ($mod ~ 2\pi$), the $SU(2)_L$ symmetry
breaks down and $W$ bosons acquire a mass $m_W$ given by
\beeq
\quad
m_W \sim \sqrt{\myfrac{2k}{\pi R}}  ~ e^{-\pi kR} 
~ \sin \myfrac{\theta_W}{2} ~.
\label{Wmass1}
\eneq
In the RS warped space the Kaluza-Klein mass scale $M_{KK}$, 
characterizing a mass spectrum $m_n \sim  n M_{KK}$  is given by
\beqn
\quad
M_{KK} &\sim& \myfrac{\pi k}{e^{\pi kR} - 1} \cr
&=& \cases{ 1/R &for $k \go 0$\cr 
\pi k e^{-\pi k R} &for $kR > 2$~.\cr}
\label{KK1}
\eeqn

In a typical model $\theta_W$ takes a value  $(0.2 \pi \sim 0.5 \pi)$.
To yield $k=M_{pl}$ in (\ref{Wmass1}), $kR$ must be $(11.75 \sim 12.0)$.
Note that $kR=6$ and 24 yield $k= 10^{11}$ GeV and $10^{36}$ GeV, respectively.
Thus the value of $kR$ is determined to be $12 \pm 0.3$.

Combining (\ref{Wmass1}) and (\ref{KK1}), one obtains 
\beeq
\quad
M_{KK} \sim \myfrac{\pi}{\sin \onehalf \theta_W} 
\bigg( \myfrac{\pi kR}{2} \bigg)^{1/2} m_W ~.
\label{KK2}
\eneq
This should be compared with the formula in flat space;
$M_{KK} \sim (2\pi/\theta_W) m_W$.  There appears an enhancement factor
$\onehalf \pi kR \sim 20$ in the RS warped  space.
Inserting the value for $kR$, one finds that
\beeq
\quad
M_{KK} = 1.6~{\rm TeV} \sim 3.6~ {\rm TeV} ~.
\label{KK3}
\eneq
At LHC,  Kaluza-Klein excited states can be produced in 
intermediate processes so that their existence can be indirectly
checked for the value in (\ref{KK3}).

\section{Higgs particles}
The Higgs field $\phi$ in four dimensions corresponds to fluctuations
of the non-Abelian Aharonov-Bohm phase $\theta_W$. More explicitly
\beeq
\quad
\theta_W \go g_4 \bigg( \myfrac{\pi kR }{2} \bigg)^{1/2} ~
\myfrac{v+ \phi}{M_{KK}}
\label{higgs1}
\eneq
where $g_4$ is the four-dimensional gauge coupling constant. 

At the quantum level the effective potential for $\theta_W$
 becomes nontrivial.  Expanding it around its
global minimum, one finds
\beqn
\quad
V_\eff (\theta_W) &=& {\rm const.} +
\myfrac{m_H^2}{2} \phi^2 \cr
&&\hskip .5cm
+ \myfrac{\eta}{3} \phi^3 
+ \myfrac{\lambda}{4} \phi^4 + \cdots ~.
\label{effV1}
\eeqn
The effective potential is determined, once the mass spectrum
$ m_n(\theta_W)$ is found for each field.
It is shown that
\beeq
\quad
V_\eff (\theta_W) = \myfrac{3}{128 \pi^6} M_{KK}^4 f(\theta_W)
\label{effV2}
\eneq
where $f(\theta_W + 2\pi) = f(\theta_W)$ is a periodic function
with an amplitude of O(1).

It follows from (\ref{effV1}) and (\ref{effV2}) that
\beeq
\quad
m_H = \sqrt{ \myfrac{3 \alpha_W}{32\pi} f^{(2)}(\theta_W) }
~ \myfrac{\pi kR}{2} \myfrac{m_W}{\sin \onehalf \theta_W}
\label{higgs2}
\eneq
and
\beeq
\quad
\lambda = \myfrac{\alpha_W^2}{16} f^{(4)}(\theta_W) 
\bigg( \myfrac{\pi k R}{2} \bigg)^2
\label{quartic1}
\eneq
where $\alpha_W = g_4^2/4\pi \sim 0.03$.  Notice the appearance of 
the enhancement factor $\onehalf \pi kR$ in (\ref{higgs2}) and (\ref{quartic1}),
which distinguishes the formulas in the warped space from those in 
flat space.  In a typical model we have found that 
$f^{(2)}(\theta_W)$ and $f^{(4)}(\theta_W)$ are  about 4.  
For $\theta_W = (0.2 \sim 0.5) \pi$, one finds 
$m_H = (125 \sim 286)$ GeV!  (The experimental bound is
$m_H > 116$ GeV.[10])  It is remarkable that
the dynamical gauge-Higgs unification predicts the mass of
the Higgs particle exactly in the range where experiments
at LHC will explore.  The quartic coupling constant $\lambda$ is
predicted to be $\sim 0.09$, though there is ambiguity in the 
value of $f^{(4)}(\theta_W)$.  

We summarize the prediction for $M_{KK}$, $m_H$, and $\lambda$ in
Table 1.  The values for $M_{KK}$ and $m_H$ predicted in flat space 
are inconsistent with the observation, but the situation drastically 
changes for the better in the Randall-Sundrum warped space.  

\vskip .3cm
\begin{table}[bh]
\begin{center}
\begin{tabular}{|c|c|c|}     \hline
\multicolumn{1}{|c|} { } &{flat space}
        & {RS space}  \\   \hline 
\multicolumn{1}{|c|} {$M_{KK}$} & {$320 \sim 800$ GeV}
        &  {$1.6 \sim 3.5$ TeV} \\ \hline
\multicolumn{1}{|c|} {$m_H$} & {$6 \sim 15$ GeV}  &
{$125 \sim 286$ GeV} \\ \hline
\multicolumn{1}{|c|} {$\lambda$} & {0.00025}  &
{0.09} \\  \hline
\end{tabular}
\end{center}
\vskip .2cm
\caption[table-higgs]{Prediction in the dynamical gauge-Higgs unification scenario
in flat space and in the Randall-Sundrum warped space.  $\theta_W$ is chosen
to be $0.2 \pi \sim 0.5 \pi$.}
\label{table-higgs}%
\end{table}


\section{Quarks and leptons}
Another magic in the RS warped space appears in the fermion sector.[11]
Each multiplet of fermions enters as a 5-d Dirac fermion in
a triplet representation of $SU(3)$.  For instance, a lepton
multiplet $\psi$ in the first generation consists of
\beeq
\quad
\psi_L= \pmatrix{\nu_L \cr e_L \cr \tilde e_L} ~,~
\psi_R= \pmatrix{\tilde \nu_R \cr \tilde e_R \cr e_R} ~.
\label{fermion1}
\eneq
The components $\nu_L$, $e_L$, and $e_R$ have zero modes which
appear as 4-d $\nu_L$, $e_L$, and $e_R$.  On the other hand
$\tilde \nu_R$, $\tilde e_R$, and $\tilde e_L$ have no zero 
modes so that they drop from the particle spectrum in four
dimensions at low energies.

The Lagrangian density for a fermion multiplet in the RS space 
is given by
\beqn
{\mathcal L} &=& \psibar i \Gamma^a {e_a}^M D_M \psi 
  - c\, k \, \psibar \psi ~, \cr
\noalign{\kern 5pt}
\quad
D_M &=& \dd_M + \frac{1}{8} \omega_{bcM} [\Gamma^b,  \Gamma^c] - ig A_M ~.
\label{fermion2}
\eeqn
The covariant derivative $D_M$ is dictated by the general coordinate
invariance and the gauge invariance.  $ck \psibar \psi$ is 
called the bulk kink mass term.[12]  $c$ is a dimensionless
parameter.

Although $c$ is called a bulk mass parameter, quarks and leptons 
remain massless even with $c \not= 0$ 
unless the electroweak symmetry breaks down.  
Their wave functions in the fifth dimension, however, depend on the 
value $c$.  When the electroweak symmetry breaks down with 
nonvanishing $\theta_W$, those quarks and leptons acquire finite
masses given by
\beeq
m_f = \sqrt{ \myfrac{\pi kR}{2} 
\myfrac{(1-4c^2)(z_1^2 -1)}{(z_1^{1-2c} -1)(z_1^{1+2c} -1)} } ~  m_W
\label{fermion3}
\eneq
where $z_1 = e^{\pi kR}$.  A fermion mass is determined by $c$, 
or vice versa.  See fig.\ 2.

$c = \pm \onehalf$ corresponds to $m_f = m_W$.  Except for top quarks
all fermions have $|c| > \onehalf$.  As shown in Table 2, 
the top quark mass corresponds to $c=0.43$ whereas the electron
mass to $c=0.87$.  Although $m_t/m_e \sim 10^5$, there appears 
no hierarchical structure in the $c$ space.  This gives us 
a good hint to understand the hierarchy in the quark-lepton mass spectrum.

\begin{figure}[hbt]
\begin{center}
\includegraphics[width=8.cm]{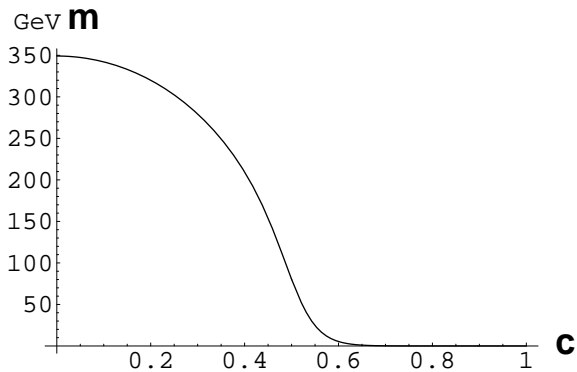}
\caption{The bulk mass parameter $c$ v.s. the fermion mass $m_f$
in (\ref{fermion3}).}
\end{center}
\end{figure}

\begin{table}[hb]
\begin{center}
\begin{tabular}{|c|c|c|}     \hline
\multicolumn{1}{|c|} {particle} &{mass (GeV)}
        & {\qquad $c$ \qquad}  \\   \hline 
\multicolumn{1}{|c|} {$e$} & {$0.510 \times 10^{-3}$}
        &  {$0.87$} \\ \hline
\multicolumn{1}{|c|} {$\mu$} & {$0.105$}
        &  {$0.72$} \\ \hline
\multicolumn{1}{|c|} {$\tau$} & {$1.777$}
        &  {$0.63$} \\ \hline
\multicolumn{1}{|c|} {$u$} & {$4 \times 10^{-3}$}
        &  {$0.81$} \\ \hline
\multicolumn{1}{|c|} {$c$} & {$1.3$}
        &  {$0.64$} \\ \hline
\multicolumn{1}{|c|} {$t$} & {$175$}
        &  {$0.43$} \\ \hline
\end{tabular}
\end{center}
\vskip .2cm
\caption[fermionmass]{The bulk kink mass parameter $c$ for each quark or 
lepton, following from (\ref{fermion3})}
\label{fermionmass}%
\end{table}

\section{Outlook}
The results obtained in the dynamical gauge-Higgs unification in the
Randall-Sundrum warped spacetime are surprising.  The mass of the
Higgs particle is predicted in the range 125 GeV to 285 GeV.  We have
determined the fermion wave functions in terms of their masses,
with which couplings of quarks and leptons to the KK excited states
of $W$ bosons etc.\ can be determined.   
In the  LHC experiments we might be able to see 
the trace of the extra dimension, directly or indirectly.

\section{Acknowledgement}
This work was financially supported by the Japanese Ministry of
Education and the 21st Century COE Program at Osaka University, 
{\it ``Towards a New Basic Science: Depth and Synthesis''}.

\def\jnl#1#2#3#4{{#1}{#2} (#4) #3}

\def\Zphys{{\em Z.\ Phys.} }
\def\jssc{{\em J.\ Solid State Chem.\ }}
\def\jpsJ{{\em J.\ Phys.\ Soc.\ Japan }}
\def\ptps{{\em Prog.\ Theoret.\ Phys.\ Suppl.\ }}
\def\PTP{{\em Prog.\ Theoret.\ Phys.\  }}

\def\JMP{{\em J. Math.\ Phys.} }
\def\NPB{{\em Nucl.\ Phys.} B}
\def\NP{{\em Nucl.\ Phys.} }
\def\PLB{{\em Phys.\ Lett.} B}
\def\PL{{\em Phys.\ Lett.} }
\def\PRL{\em Phys.\ Rev.\ Lett. }
\def\PRB{{\em Phys.\ Rev.} B}
\def\PRD{{\em Phys.\ Rev.} D}
\def\PRe{{\em Phys.\ Rep.} }
\def\AP{{\em Ann.\ Phys.\ (N.Y.)} }
\def\RMP{{\em Rev.\ Mod.\ Phys.} }
\def\ZPC{{\em Z.\ Phys.} C}
\def\SCI{\em Science}
\def\CMP{\em Comm.\ Math.\ Phys. }
\def\MPLA{{\em Mod.\ Phys.\ Lett.} A}
\def\IJMPA{{\em Int.\ J.\ Mod.\ Phys.} A}
\def\IJMPB{{\em Int.\ J.\ Mod.\ Phys.} B}
\def\EPJC{{\em Eur.\ Phys.\ J.} C}
\def\PR{{\em Phys.\ Rev.} }
\def\JHEP{{\em JHEP} }
\def\cmp{{\em Com.\ Math.\ Phys.}}
\def\JPA{{\em J.\  Phys.} A}
\def\JPG{{\em J.\  Phys.} G}
\def\NJP{{\em New.\ J.\  Phys.} }
\def\CQG{\em Class.\ Quant.\ Grav. }
\def\ATMP{{\em Adv.\ Theoret.\ Math.\ Phys.} }
\def\ibid{{\em ibid.} }


\end{document}